\documentclass[english,12pt]{article}
          \usepackage{babel}
          \usepackage[T1]{fontenc}
          \usepackage[latin2]{inputenc}
          \usepackage{pslatex}
\begin{document}

\title{\bf Curious Variables Experiment (CURVE). \\
CCD Photometry of Dwarf Nova V660 Herculis}
\author{A. ~O~l~e~c~h$^1$, ~K. ~Z~{\l}~o~c~z~e~w~s~k~i$^2$, 
~L.M. ~C~o~o~k$^3$,\\ ~K. ~M~u~l~a~r~c~z~y~k$^2$, 
~P. ~K~\c{e}~d~z~i~e~r~s~k~i$^2$ ~and~ M. ~W~i~{\'s}~n~i~e~w~s~k~i$^1$}
\date{$^1$ Nicolaus Copernicus Astronomical Center,
Polish Academy of Sciences,
ul.~Bartycka~18, 00-716~Warszawa, Poland\\
{\tt e-mail: (olech,mwisniew)@camk.edu.pl}\\
~\\
$^2$ Warsaw University Observatory, Al. Ujazdowskie 4, 00-476 Warszawa,
Poland\\ {\tt e-mail: (kzlocz,kmularcz,pkedzier)@astrouw.edu.pl}\\
~\\
$^3$ Center for Backyard Astrophysics (Concord), \\ 1730 Helix Court, Concord,
CA 94518, USA \\ {\tt e-mail: lcoo@yahoo.com}}
\maketitle

\begin{abstract} 

We report extensive photometry of the dwarf nova V660 Herculis. During our
campaign, lasting from August 2003 to November 2004, we recorded one
bright eruption which turned out to be a superoutburst lasting about 15
days and having amplitude of $\sim4.5$ mag. Clear superhumps with a
mean period of $P_{\rm sh} = 0.080924(18)$ days ($116.53\pm0.03$ min) were
present during all nights of the superoutburst. The period of the superhumps
was not stable and in the interval covered by our observations it
decreased with a rate of $\dot P/P_{\rm sh} = -4.0(1.4) \times 10^{-5}$. 

Basing on our data and the known orbital period of the binary
(Thorstensen and Fenton 2003) we calculate the period excess of
$3.4\pm0.1$\%, which is typical for an SU UMa star at this orbital period.
This value indicates that the mass ratio of the system is $q=0.154$.

\end{abstract}

\section{Introduction}

The first photometry of V660 Her reported in the literature was obtained
by Shugarov (1975) who observed the star in two outbursts in 1973 and 1974.
He estimated the photographic magnitude of the variable at maximum light
as $m_{\rm pg}\approx 16$.

In July 1995, V660 Her was observed during a long and bright outburst by 
Spogli et al. (1998). The observed maximum was estimated to have been
$B=14.3$ mag and the mean color indices were $(B-V)=-0.04$,
$(V-R_C)=0.06$ and $(V-I_C)=0.07$. The light curve showed a linear decay
with a rate of $0.13-0.15$ mag/day depending on the filter used.

A spectrum obtained by Liu et al. (1999) showed features typical for
dwarf novae in minimum light.

In June and July 2001 Thorstensen and Fenton (2003) obtained 58 spectra
of V660 Her in minimum light. The spectra showed single-peaked Balmer
lines which suggested a relatively low orbital inclination. The
continuum indicated $V\approx 18.7$ mag. A fit to the radial velocities
derived from spectra of V660 Her gave $K=39\pm5$ km/s and an orbital period
of $P_{\rm orb}=0.07826(8)$ days ($112.60\pm0.12$ min). The period being
shorter than two hours and relatively large amplitude of the long
outburst observed by Spogli et al. (1998) indicated that V660 Her might
be a member of SU UMa class of dwarf novae.

Recently, Kato et al. (2003) in the discussion of outburst parameters
and superhump regrowth phenomenon in SU UMa stars presented unpublished
data of V660 Her from its 1995 and 1999 outburst indicating the presence
of superhumps with a period of around 0.081 days.

\vspace{9.7cm}

\includegraphics{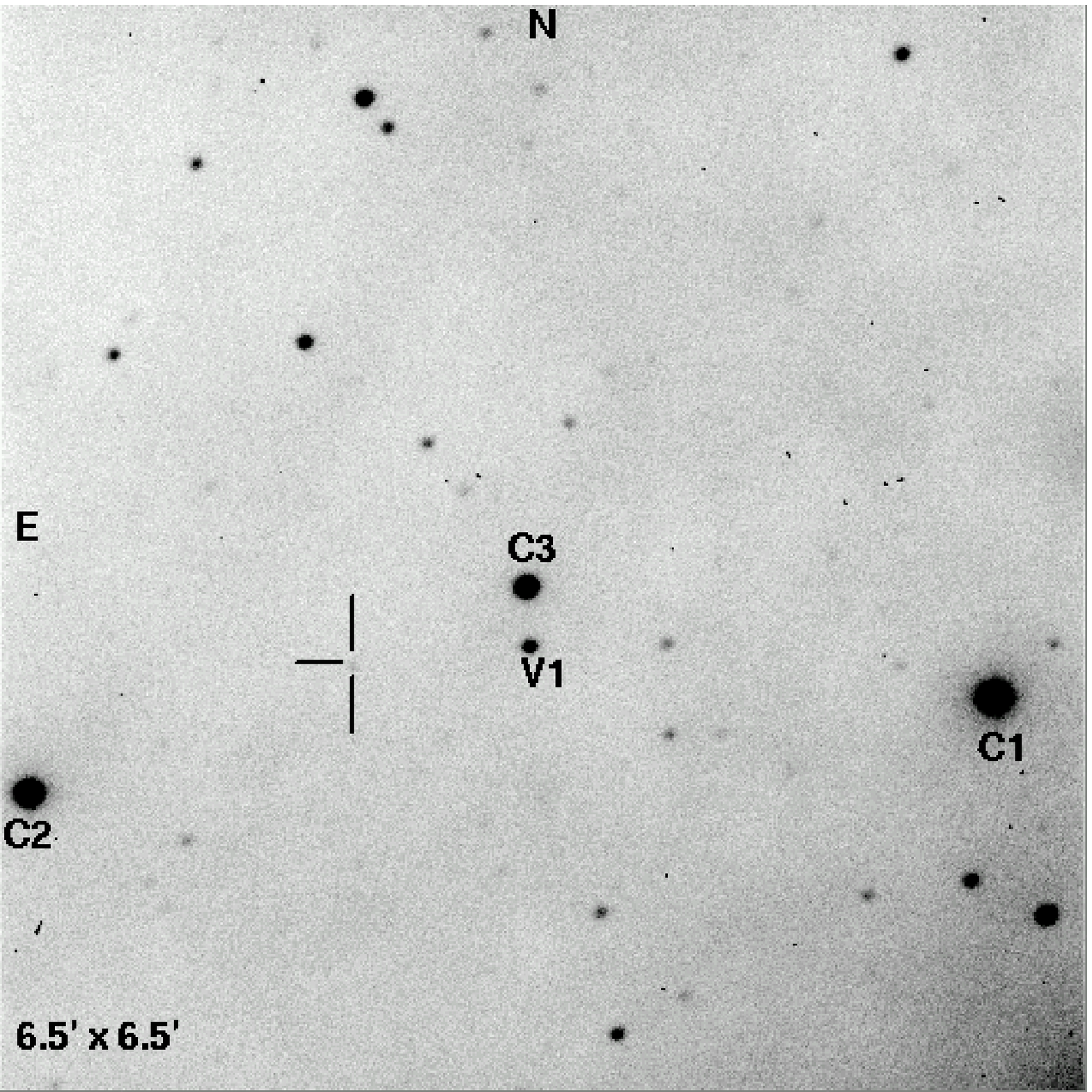}

   \begin{figure}[h]
      \caption{\sf Finding chart for V660 Her covering a region of $6.5
\times 6.5$ arcminutes. The cross matches the position of V660 Her. The
positions of the comparison stars (C1, C2, C3) and newly discovered 
variable (V1) are also shown. North is up, East to the left.}
   \end{figure}

\begin{table}[!h]
\caption{\sc Journal of the CCD observations of V660 Her.}
\vspace{0.1cm}
\begin{center}
\begin{tabular}{|l|c|c|r|r|l|}
\hline
\hline
Date & Start & End & Length & No. of & Location \\
     & 2450000. + & 2450000. + & [hr]~ & frames & \\
\hline
2003 Aug 03/04 & 2855.34466 & 2855.45862 & 2.735 & 32 & Ostrowik\\
2003 Aug 04/05 & 2856.44848 & 2856.45775 & 0.222 &  2 & Ostrowik\\
2003 Aug 05/06 & 2857.32616 & 2857.44190 & 2.777 & 32 & Ostrowik\\
2003 Aug 06/07 & 2858.42529 & 2858.46790 & 1.023 & 13 & Ostrowik\\
2003 Aug 16/17 & 2868.31160 & 2868.36422 & 1.263 & 12 & Ostrowik\\
2003 Aug 17/18 & 2869.34714 & 2869.35696 & 0.236 &  4 & Ostrowik\\
2004 Jul 31/01 & 3218.35760 & 3218.37167 & 0.338 &  6 & Ostrowik\\
2004 Aug 04/05 & 3222.34295 & 3222.36327 & 0.488 &  9 & Ostrowik\\
2004 Sep 01/02 & 3250.35955 & 3250.36389 & 0.104 &  3 & Ostrowik\\
2004 Sep 02/03 & 3251.31885 & 3251.33603 & 0.412 &  8 & Ostrowik\\
2004 Sep 03/04 & 3252.32207 & 3252.33178 & 0.233 &  5 & Ostrowik\\
2004 Sep 06/07 & 3255.32386 & 3255.33490 & 0.265 &  6 & Ostrowik\\
2004 Sep 27/28 & 3276.25836 & 3276.26305 & 0.113 &  3 & Ostrowik\\
2004 Oct 01/02 & 3280.22695 & 3280.33218 & 2.526 & 56 & Ostrowik\\
2004 Oct 03/04 & 3282.32839 & 3282.34630 & 0.430 &  9 & Ostrowik\\
2004 Oct 04/05 & 3283.21138 & 3283.22701 & 0.375 &  7 & Ostrowik\\
2004 Oct 05/06 & 3284.21235 & 3284.31969 & 2.576 & 44 & Ostrowik\\
2004 Oct 05/06 & 3284.71440 & 3284.75190 & 0.900 & 67 & CBA Concord \\
2004 Oct 06/07 & 3285.30561 & 3285.32967 & 0.577 & 12 & Ostrowik\\
2004 Oct 06/07 & 3285.60737 & 3285.74662 & 3.342 & 82 & CBA Concord \\
2004 Oct 08/09 & 3287.60794 & 3287.66448 & 1.357 & 23 & CBA Concord \\
2004 Oct 09/10 & 3288.64789 & 3288.71976 & 1.725 & 40 & CBA Concord \\
2004 Oct 10/11 & 3289.21119 & 3289.30818 & 2.328 & 47 & Ostrowik\\
2004 Oct 10/11 & 3289.60184 & 3289.71765 & 2.779 & 61 & CBA Concord \\
2004 Oct 12/13 & 3291.21140 & 3291.30844 & 2.329 & 40 & Ostrowik\\
2004 Oct 13/14 & 3292.20030 & 3292.30960 & 2.623 & 35 & Ostrowik\\
2004 Oct 20/21 & 3299.18579 & 3299.19684 & 0.265 &  7 & Ostrowik\\
2004 Nov 06/07 & 3316.19676 & 3316.20565 & 0.213 &  4 & Ostrowik\\
2004 Nov 14/15 & 3324.15641 & 3324.16168 & 0.126 &  4 & Ostrowik\\
\hline
Total          &   --   & -- & 34.68 & 673 & \\ 
\hline
\hline
\end{tabular}
\end{center}
\end{table}

\section{Observations and Data Reduction}

Observations of V660 Her reported in this paper were collected at two
locations: the Ostrowik station of the Warsaw University Observatory and
CBA Concord in the San Francisco suburb of Concord, approximately 50 km
from East of the City. The Ostrowik data were collected using the 60-cm
Cassegrain telescope equipped with a Tektronics TK512CB back-illuminated
CCD camera. The scale of the camera was 0.76"/pixel providing a $6.5'
\times 6.5'$ field of view. The full description of the telescope and
camera was given by Udalski and Pych (1992). The CBA data were collected
using f/4.5 73-cm reflector operated at prime focus on an English
cradle mount. Images were collected with a Genesis G16 camera using a
KAF1602e chip giving a field of view of $14.3'  \times 9.5'$. Images
were reduced using {\sc AIP4WIN} software (Berry and Burnell 2000).

In Ostrowik and CBA Concord the star was monitored in "white light" in
order to be able to observe it also at minimum light of around 19 mag.
We used three comparison stars which are marked in Fig. 1 as C1, C2 and
C3. The cross matches the position of V660 Her. Symbol V1 points to
the star which turned out to be a pulsating variable of RR Lyr type.

CBA Concord exposure times were 15, 20 and 30 seconds depending upon the
brightness of the star. The Ostrowik exposure times were from 100 to 180
seconds during the bright state and from 180 to 300 seconds at minimum
light. A full journal of our CCD observations of V660 Her is given in
Table 1. In total, we monitored the star for 34.7 hours during 29 nights
and obtained 673 exposures.

All the Ostrowik data reductions were performed using a standard
procedure based on the IRAF\footnote{IRAF is distributed by the National
Optical Astronomy Observatory, which is operated by the Association of
Universities for Research in Astronomy, Inc., under a cooperative
agreement with the National Science Foundation.} package and profile
photometry was derived using the DAOphotII package (Stetson 1987). The
typical accuracy of our measurements varied between 0.008 and 0.15 mag
depending on the brightness of the object. The median value of the
photometric errors was 0.011 mag

\section{The October 2004 superoutburst}

Our observations on 2004 Oct 01/02 caught V660 Her in a very bright
state. Four nights earlier the star was at least three magnitudes
fainter. The 2.5-hour observing run from Oct 01/02 showed that a clear
tooth-shape modulation with amplitude of 0.33 mag was present in the
light curve of the star. This confirms that V660 Her belongs to the SU
UMa subclass of dwarf novae and indicates that the observed brightening was
in fact a superoutburst.

\vspace{8.6cm}

\includegraphics{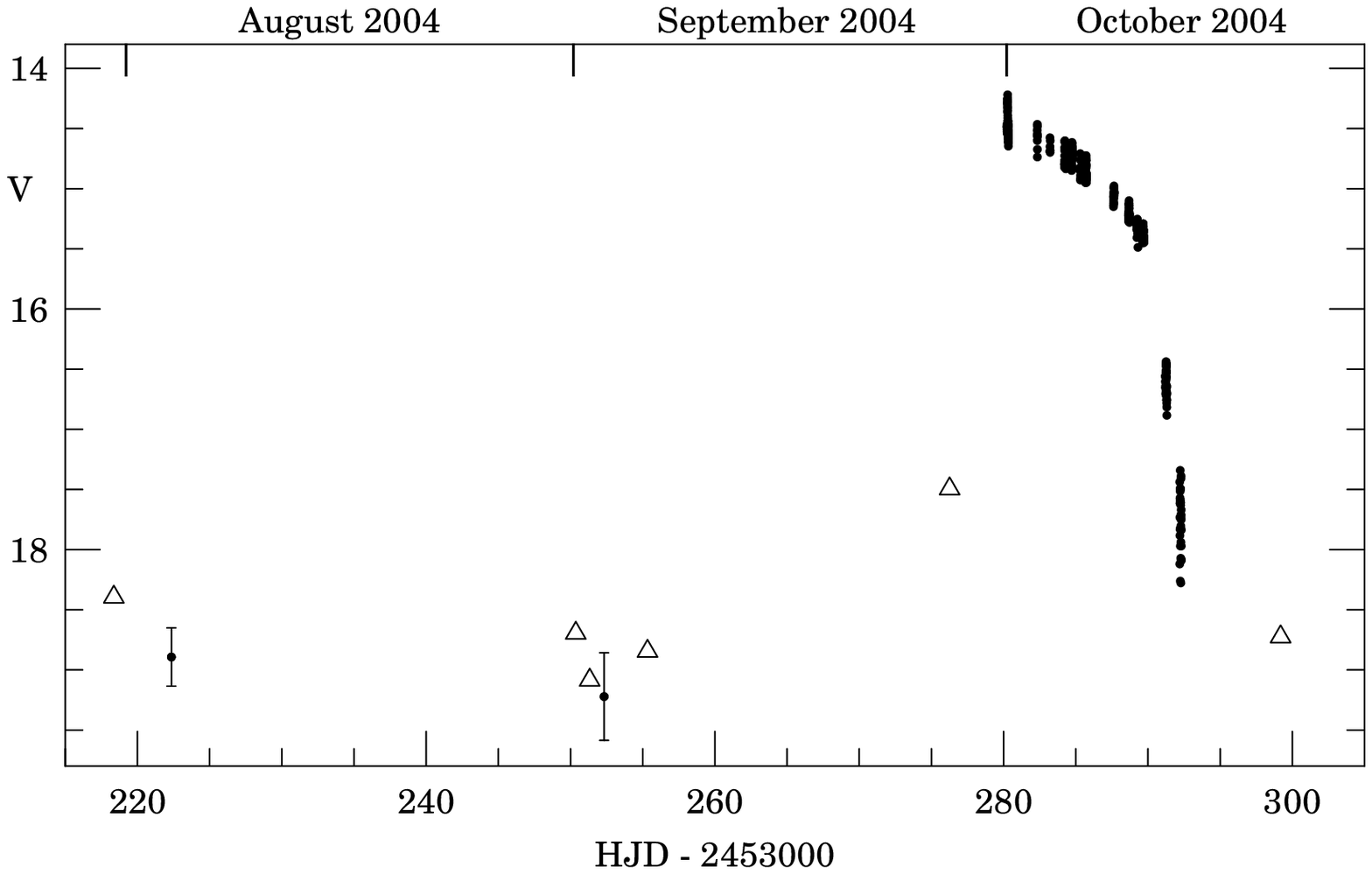}

   \begin{figure}[h]
      \caption{\sf The general photometric behavior of V660 Her during
our 2004 campaign.}
   \end{figure}

\subsection{General light curve}

Our instrumental general light curve of V660 Her was transformed to
the Johnson $V$ magnitude using data on our comparison star C2 derived by
Spogli et al. (1998). They obtained that magnitudes of this star are
$V=11.79\pm0.07$ and $R_C=11.44\pm0.05$. We additionally assumed that our
"white light" magnitudes roughly correspond to Cousins $R_C$ band
(Udalski and Pych 1992). Finally, knowing the $V-R_C$ color of V660 Her,
which was also determined by Spogli et al. (1998), we were able to
transform our light curve to Johnson $V$ band. The result is shown in
Fig. 2 where we plot the light curve of V660 Her from 
July-to October 2004. Points correspond to the magnitude determinations of
the star and open triangles denote upper magnitude limits.

During the first night of superoutburst V660 Her reached a mean magnitude
of $V=14.4$. Starting from this moment the brightness of the star was
declining with a rate of 0.11 mag/day. This "plateau" phase finished on
Oct 11 when the star entered the final decline stage and the brightness
started to decrease with a rate of 0.78 mag/day. Around Oct 14, V660 Her
returned to its quiescent magnitude of around 19 mag. The entire
superoutburst lasted of around 13-15 days.

It is interesting that during our whole campaign we did not observe any
normal outbursts. The characteristics of the eruption recorded by Spogli
et al. (1998) in July 1995 suggest that they also caught the star in
superoutburst. Their eruption reached $V\approx14.3$ mag and lasted at
least 15 days, which is almost the same as in our case. 

\vspace{12.1cm}

\includegraphics{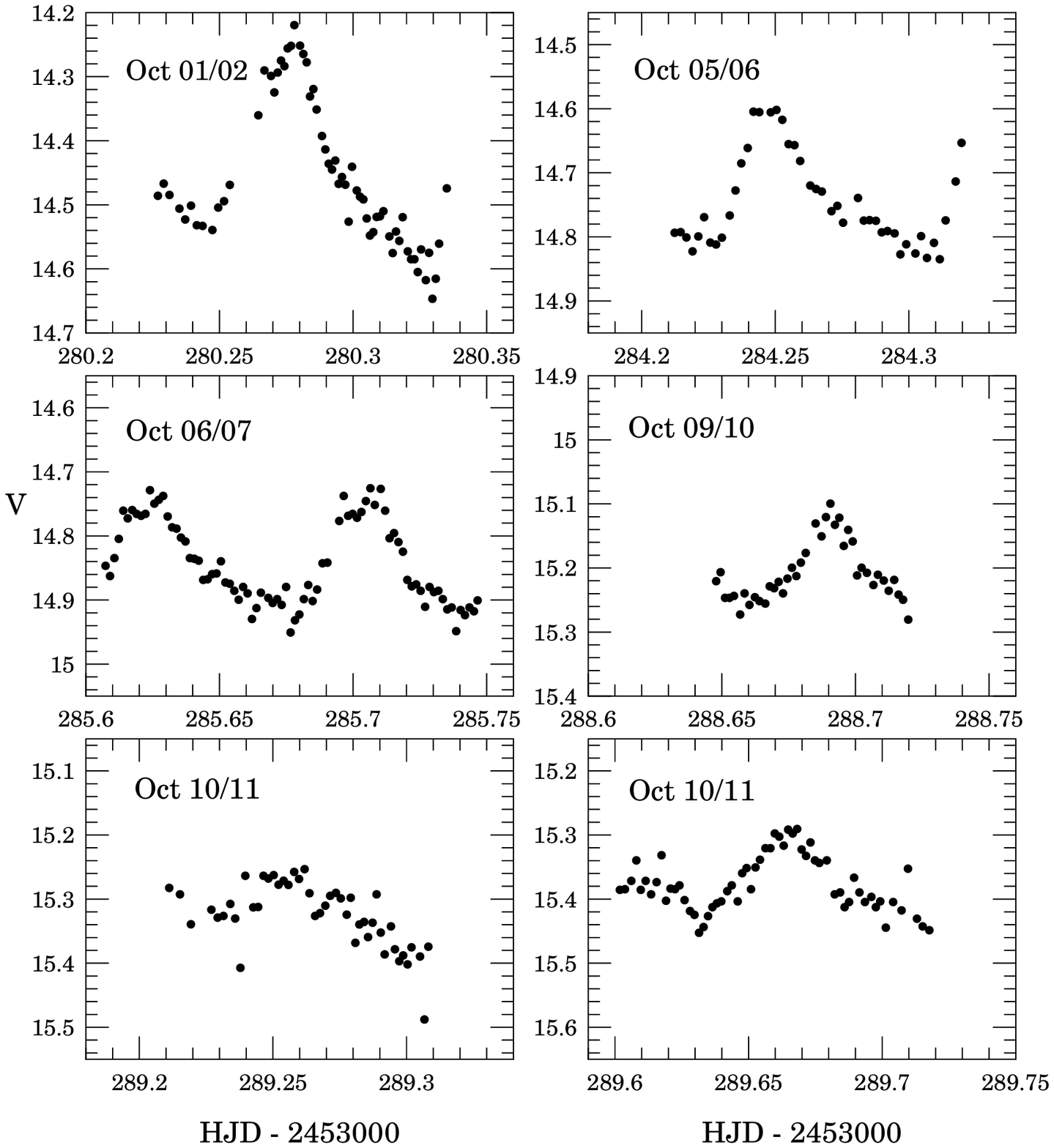}

   \begin{figure}[h]
      \caption{\sf The light curves of V660 Her during its 2004 October
superoutburst.
              }
   \end{figure}

\subsection{ANOVA analysis}

Clear superhumps were present in the light curve of V660 Her from Oct 01
to Oct 11. Sample of these superhumps is shown in Fig. 3.

From each light curve of V660 Her in superoutburst we removed the longer
term declining trend with a first or second order polynomial and
analyzed them using {\sc anova} statistics with one or two harmonic
Fourier series (Schwarzenberg-Czerny 1996). The resulting periodogram is
shown in  Fig. 4. The most prominent peak is found at a frequency of
$f_{\rm sh}=12.3573\pm0.0050$ c/d, which corresponds to a period of
$P_{\rm sh}=0.080924(33)$ days ($116.53\pm0.05$ min). 

\vspace{8.8cm}

\includegraphics{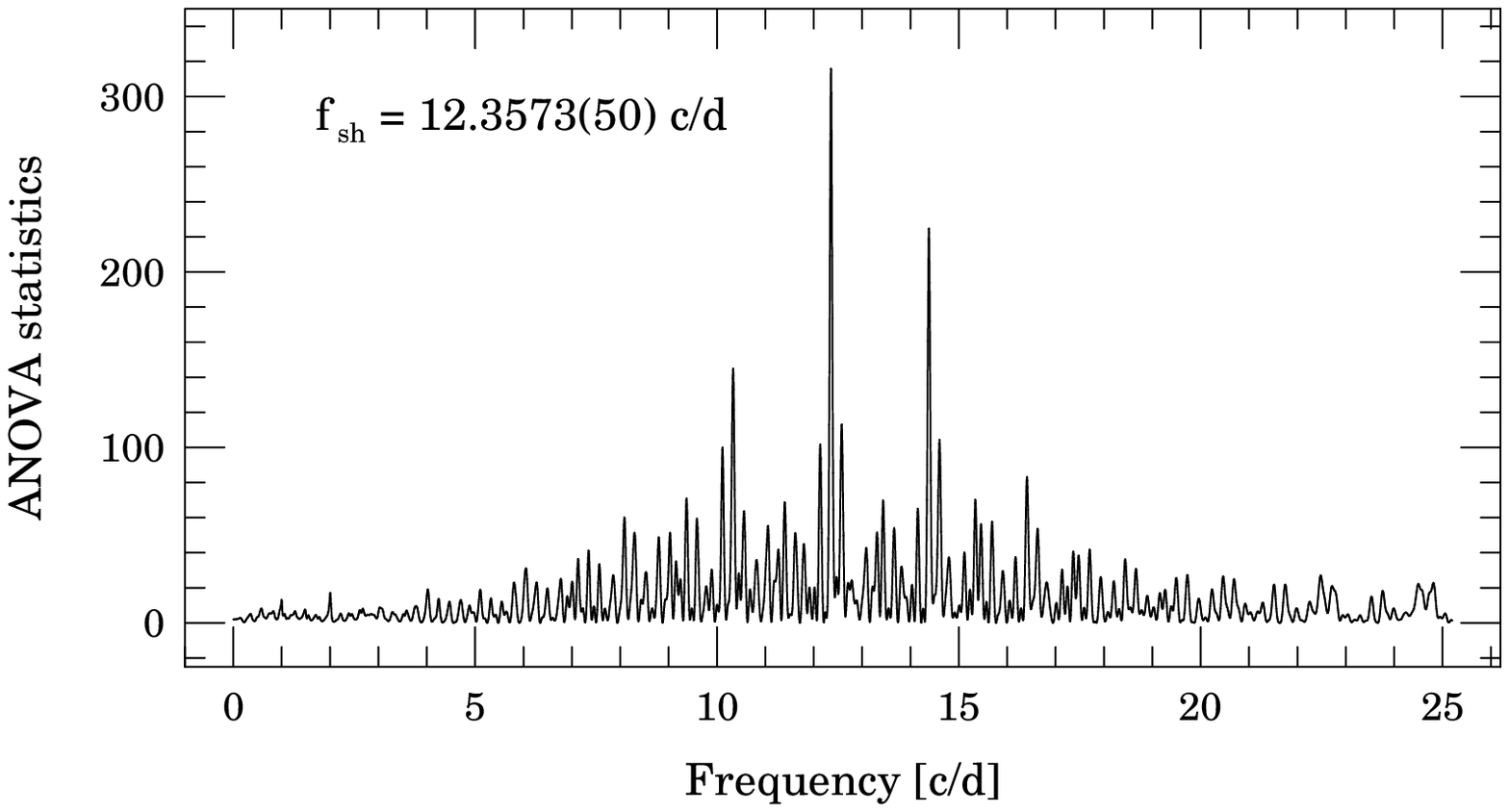}

   \begin{figure}[h]
      \caption{\sf The ANOVA power spectrum of the light curves of V660
Her from its 2004 October superoutburst.
              }
   \end{figure}

\subsection{The $O-C$ analysis}

To check the stability of the superhump period and to determine its
value in another way we constructed an $O-C$ diagram. We decided to use
the timings of primary maxima, because they were almost always more
clearly visible than minima. In the end, we were able to determine 9
times of maxima and they are listed in Table 2 together with their
errors, cycle numbers $E$ and $O-C$ values.

\begin{table}[!h]
\caption{\sc Times of maxima in the light curve of V660 Her during its
2004 October superoutburst.}
\vspace{0.1cm}
\begin{center}
\begin{tabular}{|c|c|c|r|}
\hline
\hline
Cycle & $HJD_{\rm max}-2453000$ & Error & $O-C$ \\
number $E$ & & & [cycles] \\
\hline
0 & 280.2786 & 0.0015 & $-0.021$ \\
49 & 284.2468 & 0.0020 & $0.015$ \\
55 & 284.7330 & 0.0025 & $0.023$ \\
66 & 285.6235 & 0.0025 & $0.027$ \\
67 & 285.7060 & 0.0025 & $0.047$ \\
91 & 287.6473 & 0.0030 & $0.036$ \\
104 & 288.6909 & 0.0030 & $-0.068$ \\
111 & 289.2583 & 0.0035 & $-0.056$ \\
116 & 289.6652 & 0.0035 & $-0.028$ \\
\hline
\hline
\end{tabular}
\end{center}
\end{table}

The least squares linear fit to the data from Table 2 gives the
following ephemeris for the maxima:

\begin{equation}
{\rm HJD}_{\rm max} =  2453280.2803(14) + 0.080924(21) \cdot E
\end{equation}

The $O-C$ values computed according to the ephemeris (1) are listed in
Table 2 and also shown in Fig. 5. It is clear that superhump period
shows slight decreasing trend. It is confirmed by the second order polynomial
fit to the moments of maxima which gives the following ephemeris:

\begin{equation}
{\rm HJD}_{\rm max} =  2453280.2785(15) +  0.081089(62)\cdot E - 1.61(57)
\cdot 10^{-6} \cdot E^2
\end{equation}

Finally, we can conclude that period of the superhumps was not stable
during October 2004 superoutburst of V660 Her and in the interval
covered by our observations it was decreasing with a rate of $\dot
P/P_{\rm sh} = -4.0(1.4) \times 10^{-5}$.

\vspace{9cm}

\includegraphics{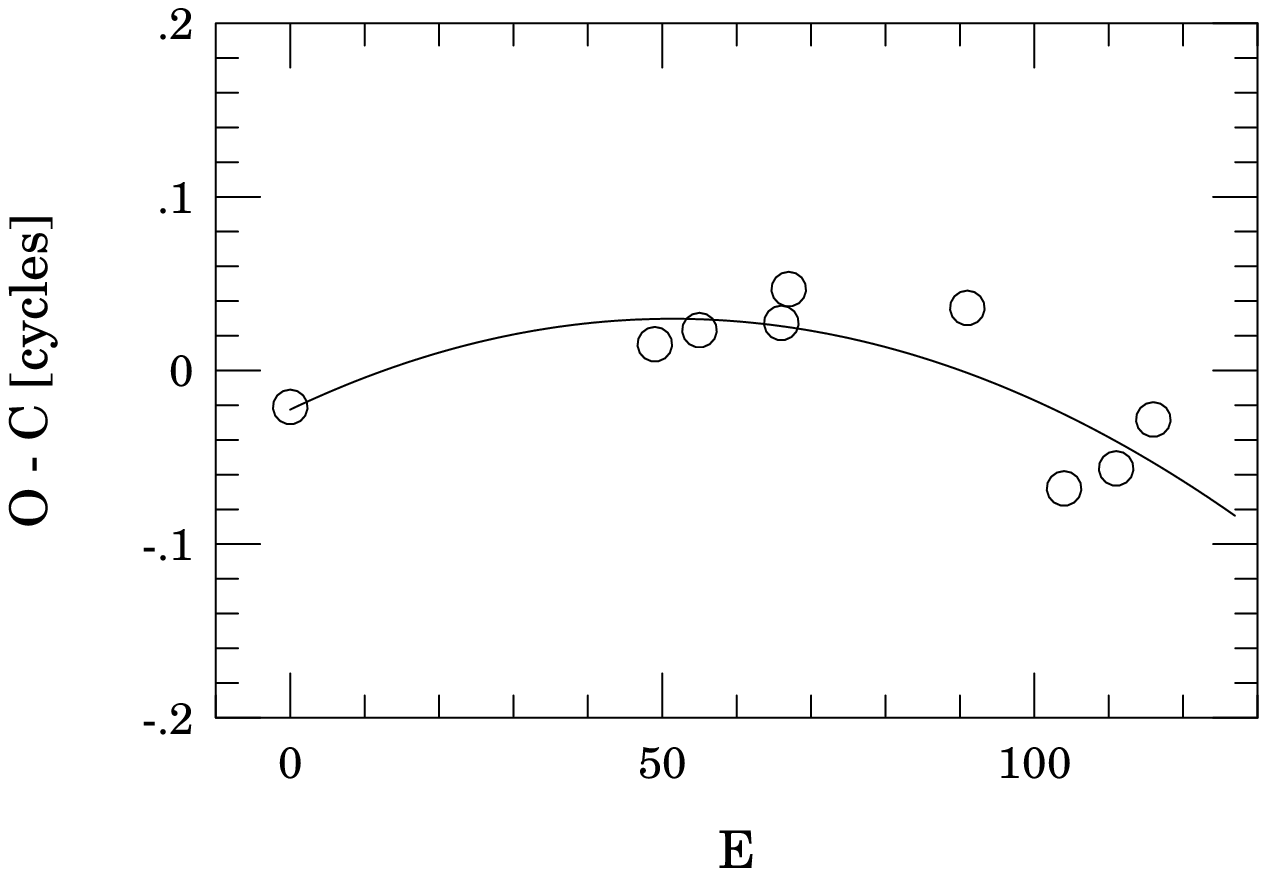}

   \begin{figure}[h]
      \caption{\sf The $O-C$ diagram for superhumps maxima of V660 Her
detected during its 2004 October superoutburst. The solid line corresponds
to the fit given by ephemeris (2).
              }
   \end{figure}
\bigskip

Combining both of our superhump period determinations returned its mean
value which is equal to $P_{\rm sh} = 0.080924(18)$ days ($116.53\pm0.03$
min).

\section{Quiescence}

In August 2003 we observed the star during six nights in quiescence. The
runs were long enough  to find orbital modulation in the light
curves. The orbital period of V660 Her is known and its value equal to
$P_{\rm orb}=0.07826(8)$ day was determined by Thorstensen and Fenton
(2003). Looking at the light curve of V660 Her obtained on 2003 Aug
03/04 and shown in Fig. 6 one can clearly see the wave in which maxima are
separated by about 0.078 day. It means that the orbital hump is visible
in the light curve of the star in quiescence and confirms the
hypothesis of Thorstensen and Fenton (2003) that the binary has a
relatively low inclination.

\vspace{9cm}

\includegraphics{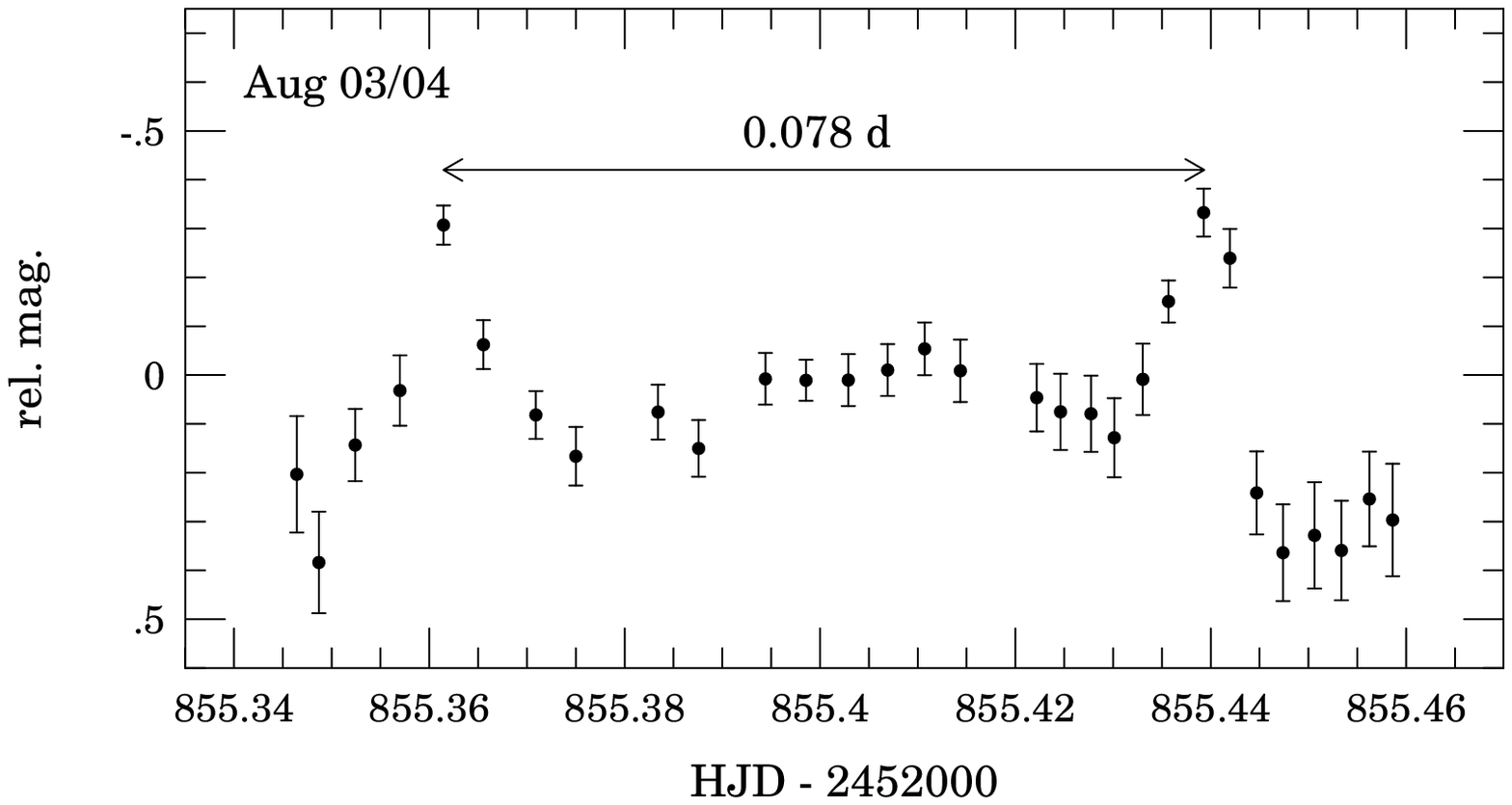}

   \begin{figure}[h]
      \caption{\sf The quiescent light curve of V660 Her from 2004 Aug 03/04.
              }
   \end{figure}
\bigskip

The ANOVA power spectrum computed for light curves from period Aug
03/04 - Aug 16/17 shows the presence of periodic or quasi-periodic signals
which are manifested by a series of peaks in frequency range from 12 to 18
c/d. Faintness of the star in quiescence and relatively short observing
runs do not allow us to make a more sophisticated analysis.

\section{Conclusions}

Superhumps - the main characteristic of SU UMa stars - occur at a period
slightly longer than the orbital period of the binary system. They are
most probably a result of accretion disc precession caused by 
gravitational perturbations from the secondary. These perturbations are
most effective when disc particles, moving on eccentric orbits, enter the
3:1 resonance. Then the superhump period is simply the beat period
between orbital and precession rate periods (Osaki 1985):

\begin{equation}
\frac{1}{P_{\rm sh}} = \frac{1}{P_{\rm orb}} - \frac{1}{P_{\rm prec}}
\end{equation}

Defining the period excess $\epsilon$ as:

\begin{equation}
\epsilon = \frac{\Delta P}{P_{orb}} = \frac{P_{sh} - P_{orb}}{P_{orb}}
\end{equation}

\noindent simple calculations lead us to the following relation
between period excess and mass ratio of the binary $q=M_2/M_1$:

\begin{equation}
\epsilon\approx\frac{0.23q}{1+0.27q}
\end{equation}
\medskip

From the observational point of view, it was first noticed by Stolz \&
Schoembs (1984) that $\epsilon$ grows with $P_{orb}$. Fig. 7 shows the
Stolz \& Schoembs relation for ordinary SU UMa-type stars (open circles)
and objects other than dwarf novae (filled circles). Data for this graph
are taken from Patterson (1998), Olech et al (2003a, 2004a) and the
newest alerts published in the VSNET archive\footnote{For clarity in
Fig. 7 we do not show the recent result of Smak and Waagen (2004) who
detected the superhumps in 1985 bright outburst of U Gem, which orbital
period is 254.7 min and period excess ~$\epsilon = 0.130 \pm 0.014$}. 
V660 Her is plotted with a filled triangle. The period excess of the
star is $\epsilon=0.034\pm0.001$ and according to the equation (5) this
corresponds to the mass ratio $q=0.154$.

\vspace{9cm}

\includegraphics{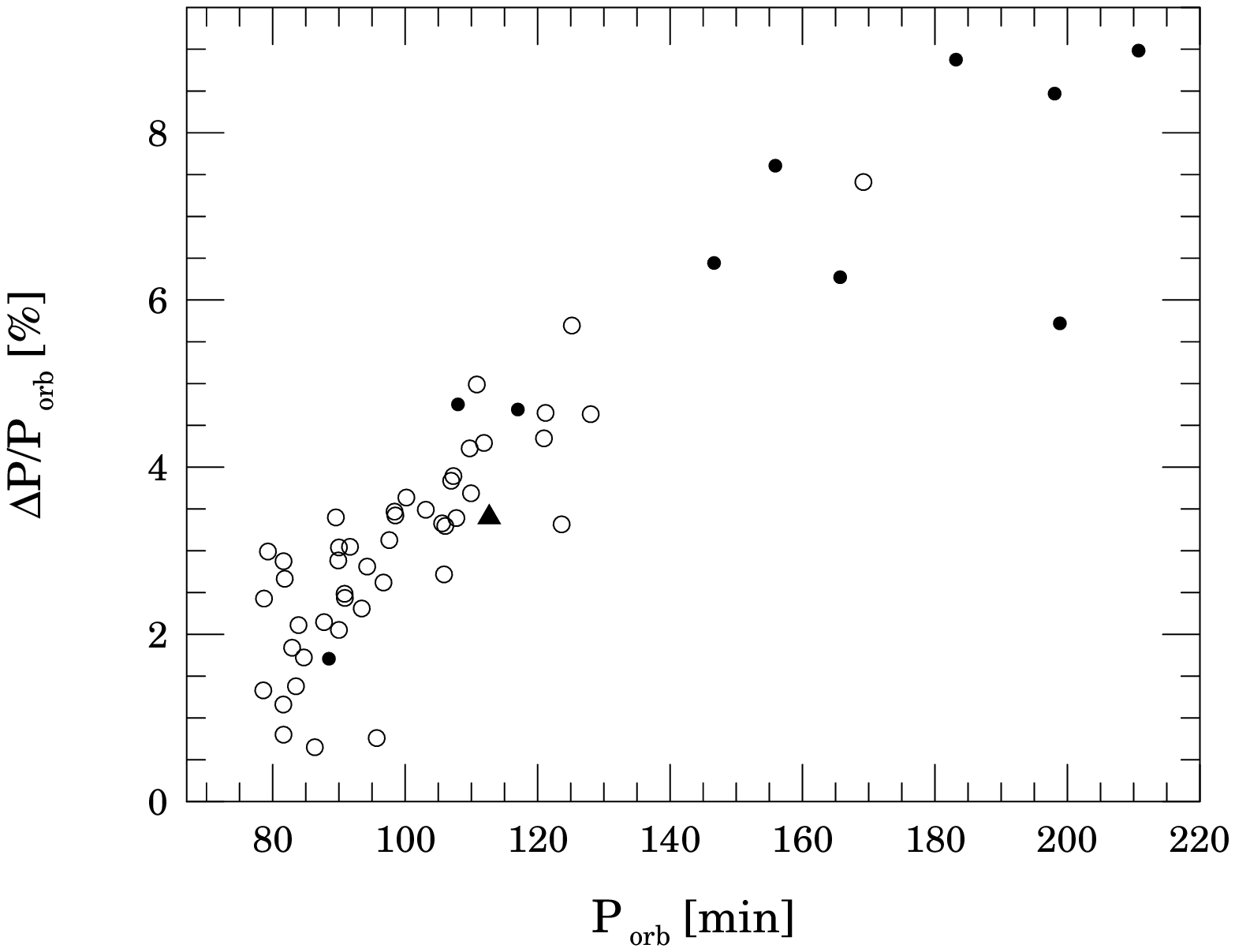}

   \begin{figure}[h]
      \caption{\sf Stolz and Schoembs relation for ordinary SU UMa
stars (open circles) and stars other than dwarf novae (filled circles).
The position of V660 Her is denoted by the solid triangle.
              }
   \end{figure}
\bigskip

Until the mid of 1990's all members of the SU UMa group seemed to show
only negative superhump period derivatives (Warner 1995). This was
interpreted as a result of disk shrinkage during the superoutburst, thus
lengthening its precession rate (Lubow 1992). This picture became more
complicated in 1997 when the first stars with $\dot P>0$ were discovered
(Semeniuk et al. 1997, Olech 1997). Presently, we know even more complex
behaviour as was shown by Olech et al (2003b, 2004b) who investigated
the $O-C$ diagrams for KS UMa, ER UMa, V1159 Ori, CY UMa, V1028 Cyg, RZ
Sge, SX LMi and TT Boo and claimed that most (probably almost all) SU
UMa stars show decreasing superhump periods at the beginning and the end
of superoutburst but increasing period in the middle phase. Recently,
Uemura et al. (2005) suggested that superhump period change might be
connected with a presence of the precursor in the light curve of
superoutburst.

Our observational coverage of superoutburst of V660 Her is far from
completeness especially in the first phase.  The observed superhump
period change equal to $\dot P/P_{\rm sh} = -4.0(1.4) \times 10^{-5}$ 
is typical for the last and longest stage of the superoutburst. Thus the
first two phases, in which the superhump period derivatives are at the
beginning negative and further positive, could be not covered by our
observations.

\bigskip \noindent {\bf Acknowledgments.} ~We acknowledge generous
allocation of  the Warsaw Observatory 0.6-m telescope time. Data from
AAVSO and VSNET observers are appreciated. This work was supported
by KBN grant number 1~P03D~006~27.

\end{document}